# A calendar *Quipu* of the early 17th century and its relationship with the Inca astronomy.

**Laura Laurencich-Minelli**
Dipartimento di Paleografia e Medievistica, Università di Bologna
Piazza S. Giovanni in Monte 2
40124 Bologna

**Giulio Magli**
Dipartimento di Matematica del Politecnico di Milano
Piazza Leonardo da Vinci 32
20133 Milano

## 1. Introduction: the Miccinelli Documents

The so-called *Miccinelli documents* are two secret Jesuit manuscripts: *Exsul Immeritus Blas Valera Populo Suo* [The unjustly banished Blas Valera to his people], (here referred as EI) and *Historia et Rudimenta Linguae Piruanorum* [History and elements of the Peruvian language], (here referred as HR). The document EI, dated Alcalà de Henares (Spain) May 10th, 1618, was written and signed by Blas Valera, a Jesuit scholar which was known prior to the discovery of the documents because his works are cited as erudite sources in the chronicle written by Garcilaso de la Vega. The document HR is instead a collection of writings, composed in Peru between 1600 and 1638 by the Italian Jesuits Johan Antonius Cumis and Johan Anellus Oliva (the second one was already known as the writer of a chronicle, which did not obtain the *imprimatur* by the Company). The two documents were presented to scholars between 1996 and 2001 and they have been recently made available to scholars with an exhaustive publication[1]; it is worth noticing that their authenticity has been proved beyond any possible doubt.[2] The aim of the present paper is to discuss in full details a calendrical document which is attached to EI; in order to focus better on this document, we recall here briefly the contents of EI and HR.

Both the manuscripts deal with the surprising personal history of the mestice Blas Valera. He was a very erudite Jesuit but his positions were very critical about the destruction of the natives and of their culture due to the Spanish invaders. As a consequence, according to what both documents say, the F. General of the Jesuit Order Claudio Acquaviva exiled him into Spain (1592). However, Valera kept on spreading his critics on the Spanish invasion and on the resulting evangelization of Perù which was carried out without respecting the culture of the natives. As a consequence, Acquaviva imposed to him a "juridical " death. Before this fictitious death, according to the documents, F. Valera gave to Garcilaso de la Vega his personal work on the history of the Inca, asking him to diffuse it respecting the Author; but Garcilaso wrote the *Comentarios Reales* twisting Valera's ideas with the aim of presenting the Incas as savages, in order to comply with the Conquerors' power; further, he acknowledged Valera's authorship in a very incomplete manner. However, according to both documents EI and HR, F. Muzio Vitelleschi helped F. Blas to go back to Perù (1598) and let him write the defense of the natives called *Nueva Coronica y Buen Gobierno* (here referred to as NC) that Valera wrote concealing himself under the name of the native Guaman

Poma de Ayala. On 1618, the same F. Muzio Vitelleschi, at that time General of the Order, helped Valera to return to Spain, in Alcalà de Henares, where he died in 1619.

The detailed study of the three documents NC, EI and HR, reveals that their aim is the realization of a kind of Inca Christian state in the Jesuitical *Provincia Peruviana*, within the Spanish kingdom. The NC is indeed a wide "letter", addressed to the king, which proposes the foundation of an Inca-Christian realm submitted to the Spanish crown in order to save the natives from the destruction. It is signed by the native Guaman Poma de Ayala, but Valera in EI as well as Anello Oliva in HR credit Blas Valera as the real mind who conceived it, and furnish convincing proofs of this fact. Besides, Oliva in HR gives the key for reading the symbols (named *tocapus*, see Sec. 2 below) which are painted in NC, claiming that they contain some messages for natives.

The ms. EI, written in Alcalà de Henares and signed "Blas Valera", is a bi-cultural document written in Latin in Europian characters and in Quechua by using traditional writings (Quipus, *tocapus* and pictography). The parts written in Latin are dedicated to F. Muzio Vitelleschi and contain claims for the respect of the Indios as persons and the restitution of their lands in the name of Christian fraternity, in order to establish a kind of Inca-*reduccion* within the *Provincia Peruviana*. With this goal in mind, efforts are made to describe the religion of the Incas as the product of the *amautas*, (a sort of philosophers) putting forward a supposed similarity between the Inca culture and the new Church of the origins in Perù. In this way, the author defends the introduction of Christianity among Peruvians trough discussions - i.e. Natural Philosophy - rather than with fighting and destroying the pre-existing religion and culture. The feasibility of such a project - at least in the mind of Blas Valera- appears to be based on the claim that the Inca religion was only in appearance a polytheist religion, because its gods and cosmic forces could be expressed in terms of "sacred numbers" which added themselves to form an unique "Divinity". This concept is synthesized in the sacred quechua chant called *Sumac Nusta*, which is reproduced as a *capacquipu* signed Blas Valera in HR, and under the numerical form of a *yupana* (abacus) both in EI and in NC.[3]

The parts of EI which were written with the traditional systems are dedicated to the natives, especially the Incas, and describe the new, Christian state of the Incas at Paititi (Bolivia), viewed as the continuation towards the East, that is "towards the rising Sun", of the last Inca Reign of Vilcabamba (destroyed with the execution of Inca Tupac Amaru in Cuzco in 1572). The *Tahuantinsuyu* (the "state of the four parts", as the Inca named their empire) is hence seen as just one *suyu* (part) Reign, the Antisuyu, whose center was Paititi, located at the confluence of two great rivers (Madre de Dios and the Beni River) resembling as a "sacred geography" the former capital of the Inca, Cusco.[4]

From the point of view of the religion, EI asserts to the natives that the Christianity practiced in this new Reign Inca has to be a sort of improvement of the ancient religion, in which the Inca gods Pachacamac and Viracocha apparently play the role of Father God and of his son Jesus. From the political and administrative point of view this ideal Inca state should be a diarchy between the Inca and the King of Spain organized with the *ceque* system (system of radial lines) as was the Inca capital Cusco. The official language of the state has to be the Quechua, written with the traditional systems, i.e. *Quipus, tocapus* and *pictography*, and the passing of time has to be accounted for using the traditional Inca calendar. Clearly, however, this last requirement would have been at high risk of being considered as idolatry. As a consequence, rather than stating it explicitly, the author's choice is to present an example of such a calendar, referred to a pre-conquest "time"; actually, as we shall se later, just the "last time" of the Incas. In this kind of calendar, all the celestial events as well as the cosmic time, the *pacha*, had (and should have in Paititi) to be *projected* on the "anthropized" and "chequered" earth, *Pachamama*, which in turn is - through the *yupana* - transformed in a sort of geo-shaman table.[5] With this procedure the shaman-astronomer (*amauta*) succeeds to predict - and therefore to control - the celestial cycles[6]. The abovementioned example of such traditional calendars is enclosed in EI in the form of a drawing: it is named *Pachaquipu* (the Quipu of the time, i.e. a device made out of cords and knots) and it is the object of our study in this article.

Finally, the miscellaneous set of documents HR is an encrypted document of European type, written in Perù (in Latin and in Italian) in subsequent periods of the first half of 17$^{th}$ century by two Italian Jesuits, Antonio Cumis and Anello Oliva, in order to testimony the development of the plan of the Inca Christian state of Paititi and of the syncretism within the Christian religion. The document is concluded by a short note added in 1737 by F. Pedro de Ilanes; in the note, written in Spanish, he cautiously asserts that actually the entire plan of Paititi was put to an end because it was heretical. In any case, it turns out clearly that at the beginning of 17$^{th}$ century the Andeans practiced a form of syncretism; the concept of divinity as described by Valera confirms the intuition by Valcarcel[7] that the pre-Columbian Andean religion was in some sense monotheist but the missionaries did not recognize it, because of their preconceptions of idolatry.

## 2. Description of the *Pachaquipu*

The object of our study, the *Pachaquipu,* is a single sheet of paper, which was originally enfolded two times and inserted in EI c.18. On the *rectum*, by the same calligraphy and the same lamp-black ink of EI, the author, Blas Valera, wrote a few sentences. These sentences are a reprimand to Garcilaso de La Vega for his ignorance on the Quipus, and state that the drawing reported on the other side is a *Pachaquipu*, a "Quipu of the time" which "regards the end of the Inca empire" (supposedly the year 1532). Playing with the word "time", Blas Valera recalls also, after his fictiuos death and under a false identity, his trip back to Perù to write the *Nueva Coronica* and back again to Spain, as well as the names of the ships who let him travel back and forth. On the *versum* we find the color depiction of the *Pachaquipu,* together with two lines of writings and the initials "BV", which occur also in all the painted pages of the main document.
The drawing (Figs. 1 and 2) represents a Quipu composed by a main cord and 13 pendants. Each pendant carries a certain number of knots, for a total of 365; between some of the knots smaller cords are fastened, bringing some symbols, painted on small rectangular cartouches. From left to right we encounter twelve pendants which represent 12 sinodic months, since each cord starts with the symbol of the new moon (see below) and contains either 29 or 30 knots. The days are also counted in groups of 15 each by the use of alternate colors (red and green) and in groups of 10 by spacing the knots ten by ten. Thus, the Quipu contains a calendar of 12 months; the 1st, 3th,4th,6th,10th months are of 29 days while the others are of 30 days; there are "weeks" (periods) of 10 and of 15 days which run independently but simultaneously with the month's count, being knotted on the same pendant . The periods of ten days are marked by groups of knots, and correspond to the Inca weeks mentioned in NC (cc.255, 260) and to the decimal basis which was in use in the Quipus; the periods of 15 days are alternately marked by red and green knots and correspond to a division into two halves of the solar months of 30 days each. The two parts correspond to the periods *hanan* (red knots) and *hurin* (green knots). In the last pendant, the 13$^{th}$, there are five green knots, which represent the days needed to bring the total to 360 (due to the presence of 5 sinodic months of 29 days) and five red further knots representing additional "epagomenal" days, needed to bring the solar count to 365.
All in all, it is clear that the calendrical structure is based on 12 lunar sinodic months. Above each pendant there is a cartouche containing a symbol for the month, and over each symbol the name of the month is also written, in Quechua language but in Latin characters. Many of the symbols can be recognized as *ticcisimi* [8] which visualize the name of the months. In the description below we will put in evidence when the symbols used in the documents are also *ticcisimi* used in the main document EI, since, according to EI (add.III), they have a special sacred power that the *amauta* has to catch and order under the aspect of a *tocapu*, i.e. in the corresponding *huaca* in the Pachamama, to avoid the destroyer Amaru. In what follows we give the description and the likely interpretation of each symbol, together with a comparison of the description of each month with the corresponding month as described in the NC *calendar of feasts* (NC actually contains two calendars, see Sec. 4 for details); this comparison is definitively worth especially in view of the fact that the two documents should have been written by the same author. As a general observation it has to be noticed that the order of the months is shifted between the two sources because, as we shall show in next section, the *Pachaquipu* begins with the new moon before the winter solstice while NC *calendar of feasts* begins with the new moon after the summer solstice; notice also that the name of the months in NC have the attribute *quilla* (moon), while the present document reports *pacha* (time) (for a complete discussion of the likely reasons of such differences see Sec. 4).

For a better understanding of the months of the *Pachaquipu*, we first describe the 13th pendant (far right) and then proceed from left to right, describing the pendants 1st to 12th.

[13th pendant] *yntihuatapacyapanapacha* [time necessary for the completion of the calendar year].

The cartouche is yellow, i.e. it is the *ticcisimi ynti*, the sun. The pendant, after the 5 green knots and the 5 red-ones of the additional days, contains a directory of symbols related to astronomical events to be used for reading the calendar: we join here a very brief description of each symbol with its explication, which is written in Latin letters at the bottom of the *Pachaquipu*.

 - a square without color [the *ticcisimi quilla* = moon], the explication is: *mosocquilla* [new moon],
- a left-handed crescent, the explication is: *pacaricquilla/vinacquilla* [waxing moon],
- a white square, the explication is:  *pascaquilla* [full moon],
- a right-handed crescent, the explication is:  *huañucquilla* [waning moon],
- a yellow square [the *ticcisimi inti*= sun] covered by a bow [the ticcisimi *tuta* =night], the explication is: *hatuntuta* [Midwinter, June solstice],
- a yellow square [the *ticcisimi inti*= sun] above the bow [the *ticcisimi tuta* =night], the explication is: *hatunpunchau* [Midsummer, December solstice],
- a yellow square [the *ticcisimi inti*= sun] knotted with the bow [the *ticcisimi tuta*= night] which is on the same level as the sun, the explication is: *pituçuni* [equinoxes],
- a rectangle  1/3 white and 2/3 grey [the *ticcisimi yanpintuy* =eclipse of the moon], the explication is: *yanpintuy* [black sheet= eclipse of moon],
- a rectangle  2/3 grey and 1/3 yellow, the explication is: *yntihuañuy* [eclipse of sun],
- a rectangle containing a shining star with eight jags, the explication is: *Collcacapac* [the Pleiades]

As we shall see in next section, all these symbols are used, attached to days of the calendar, with the exception of the sun-eclipse symbol. It must also be noticed that two further symbols appear in the calendar but they are lacking both in the directory and in the explanation: a yellow C-shaped cord (which must be connected to the sun, since the yellow color is connected to the sun according to the "language of colors" discussed in EI) which holds two groups of  days, one on the 5th and the other one on the 9th pendant, and a single knot marked black, placed on the 6th pendant.

 [1° pendant] *yntiraymipacha*  =  time of the Feast of the Sun.

The cartouche contains the *ticcismi tyana* (c.5v, 6v) i.e. the throne; it is divided horizontally into two fields: the lower one is yellow, the top one is white and greater than the lower-one. The yellow field is composed horizontally by a rectangle on which it is drawn the sun-throne, i.e. a kind of saddle between two "towers" of the same height at the sides of which lie some right-angles (two on the right and one on the left). The fact that the "yellow" section is much smaller than the white is to be put in correlation to the winter solstice occurring in this month, and therefore the longest night of the year (=the biggest "space" for the moon, associated with white - compare months no. 4 and 7, containing equinox and summer solstice); perhaps the whole symbol contains also an allusion to the view at the Cusco eastern horizon, with terraced fields and the pillars that were used as "seats of the sun" to make astronomical observations (we shall encounter again such pillars later on).

In the NC (cc.246, 247) the same month is called *Huaucaicusqui quilla* (moon of the rest after the harvest) followed however by the mention of *Inti raymi,* the fest of the Sun at the winter solstice. There is therefore conceptual concordance between the two sources.

[2° pendant] *pachacyahuarllamapacha* = time of the 100 red llamas.

The cartouche represents 15 red llamas facing right, signifying a high number of such animals. The chronicler Polo de Ondegardo (I: 213) tells us that every month the Inca sacrificed 100 llamas and that those of red color where offered to the Thunder god to ask for rain.
The name of the month in the NC (cc. 248, 249) is: *Chacra ricui chacha cunacuy chaua uarqum quilla* (the moon of the inspection of the lands, of the repartition of the lands) but the Pachaquipu is in accordance both with the NC picture, representing a priest while burning an offering, and with the NC description that mentions the sacrifice of the red llamas: i.e. there is conceptual concordance between the two sources.

[3° pendant] *yapuypacha* = time to plow.

The cartouche contains the ticcisimi *tacla*, plow (EI c. 6r, 6V) that emphasizes the caption. The name of the month corresponds precisely to that of NC (cc. 250, 251) (*Chacra yapuy quilla*= the moon of plowing the fields).

[4° pendant] *coyaraymipacha* = time of the Feast of the Coya (the Moon).

The cartouche contains again the *ticcisimi tyana*, throne, quite similar to that of the first pendant; however the bottom field is white and equal in size to the top one. The fact that both fields have the color of the moon instead of that of the sun probably emphasizes that the main festivity is devoted to the moon, while the fact that the two fields are equal in size indicates the month of the equinox, when night and day have the same length and the moon and the sun are considered to have the same "strength". The name of the month is identical to that of NC cc.252, 253 (*Coya raymi quilla*= the moon of the Feast of the Coya[the Moon]).

[5° pendant] *paramañaypacha* = time to implore for the rain.

The cartouche contains a black mask with red tongue. It corresponds partially to a figure of EI; however in the latter the mask is red and grinds the teeth; the two figures might be two aspects (positive and negative respectively) of EI' *Spiritus Vitalis*, a sacred force of the fertility having a dual appearance (destroyer and fertilizer-creator). The image recalls a "monster" of the Nazca culture (Early Intermediate Period) whose body is however usually portrayed in flight.[9] (It should be noted that the use of ancient cultural elements, apparently still alive in the 16-17th century, is frequent in EI).
The name of the month in NC (cc. 254, 255) is *Uma raymi quilla* (the moon of the main feast); the text mentions a song aimed to beseech the rain to *Runa Camac*, creator, *Micocpac Runa*, builder of man, and *Vari Vira Cocha*, god of Wari. According to EI these are aspects of the same sacred *Spiritus Vitalis* which are materialized on the earth in the form of huacas, and therefore controllable by humanity. Thus, again, the two sources conceptually agree; in addition, they both allude, in a form readable only by the indigenous peoples, to a cult whose practicing was, of course, very dangerous at the times of writing of Blas Valera.

[6° pendant] *ayamarcaypacha* = time to carry the deads in a procession;

The cartouche, absent in the iconography of the main document EI, represents a red mask similar to those used on the mummies of the Central Coast (Late Intermediate Period ); thus it means that the month is devoted to the mummies. The 20[th] day of this month is evidenced by a black knot which is among the green knots, i.e. during the *hurin pacha*.
It is very interesting the fact that only the calendar of NC (cc 256, 257) mentions this month in the same way as EI Pachaquipu: *Aya Marcay quilla* , (the moon in which the [mummies of] deads are brought in procession) while more ancient sources name this month *ayarmacay*= time of the social group Ayar Maca. We agree with Zuidema[10] who noticed that *ayarmacay* might thus have been the name of this month before the 3th Limense Council (1585), and that this name was given to the month with the aim of making it more similar to our November, the month of the western cult of the deads. We also suggest that perhaps the replacement (*ayamarca* instead of *ayarmaca*) is also due to a dynastic argument: if, as it seems, the Pachaquipu should represent an example of the calendar timetable to be used in the Inca-Christian Kingdom of Paititi where all ethnic groups are unified in the figure of the Incas descendant from Paullu Inca out of the branch of Huayna Capac, then it would not make sense any more to celebrate a feast reserved for a special ethnic group which did not exist any more. Anyway the fact that there is coincidence on the name of this month only between the NC and EI calendars, suggests once more that NC and EI calendars arise from the same mind.

[7th pendant] *capacyntiraymipacha* == time of the royal feast of the Sun.

The cartouche contains the *ticcisimi tyana*, throne, in a form very similar to that of the first and of the fourth pendant; however the yellow part fills up nearly all the cartouche to indicate the great power of the sun in this month related also to the longest day of the year, while the power of the moon is low; actually one can follow the increase in the sunlight from winter solstice to summer solstice looking at the corresponding, decreasing lower section of the "throne" *ticcisimi* from the 1[st], to the 4[th] until the 7[th] pendant of the document.
In NC (cc.258, 259) the name of the 7[th] month is *Capac ynti raymi* and therefore the two sources fully coincide; this is the unique case in which the name of the month in NC is not accompanied by the lunar indication *quilla,* perhaps to emphasize the greatest power of the Sun on the moon.

[8th pendant] = *huacapacha* = Time of the huacas.

The cartouche contains the *ticcisimi* for *huaca*.
In NC (cc. 236, 237) the month is named *Capacraimicamay quilla* which is translated into "month of the rest after the great feast of capac raymi", a name apparently innocuous. However, the NC text confirms that this month was devoted to important sacrifices to the *huacas* and describes them starting from the temples of the sun and the moon, and arriving to Huanacauri and Pacaritambo. There is therefore conceptual concordance between the two sources.

[9° pendant] *huarachicuypacha* = time of the first dressing of males with perizoma;

The cartouche contains the *ticcisimi ullu* (the penis).

In NC (cc. 238. 239) the month is called *paucar uaray quilla* (the moon in which precious perizomas are worn) and the text mentions offerings to the huacas and two Inca ceremonies which are said to be forbidden: *uarachico* and *rutochico,* both associated with the life cycle. All in all, the two sources agree but the fact that the *Pachaquipu* gives the name *huarachicuypacha* to the month suggest that the *huarachicu* was the most important between the two ceremonies mentioned in NC. It is curious that only the *Pachaquipu* and the calendar of NC (cc 256, 257) name this month in this way, while other sources name it in a very different way, *Puacar huara*, and do not mention the ceremonies.

[10 ° pendant] = *paraypacha* = the time of the rain;

The cartouche portrays a personage in prayer, with long hair and wearing a black hut; the picture is new with respect to the iconography of the main document EI.
In NC (cc. 240, 241) the name of the month is *Pacha pucuy quilla*, translated as the moon of the growing in the accompanying text. The NC picture shows a cloudy sky and the Inca offering a black llama; the text mentions many ceremonies, especially sacrifices of black llamas and periods of fasting, led by sorcerers; one of them is thus probably the personage in the *Pachaquipu* cartouche of this month. The sacrifices were probably devoted to Viracocha; according to ethnographical data, the full moon before the equinox was the time of a feast devoted to the *hanan* deads, associated with the summit of the mountains. This is the unique NC month containing the term *pacha*.

[11 ° pendant] *rinrituccinapacha* == time of piercing the ears.

The cartouche shows two elongated lobes with earrings. A circular earring is depicted also in a *tocapu* of EI c. 11r with a complex caption which mentions the ears, the word, the moon, the uterus and the sperm of Pachacamac; to the same *tocapu* Oliva (HR), giving the meaning of NC *tocapus*, adds to EI a caption with the words: "evolving cycle": i.e. it means all what is evolving . If we read this caption together with the well known fact that the perforation of the ear lobes and the extension of the lobes due to heavy earrings were considered by the Inca as high status symbol, it results that piercing the ears symbolically gave to the nobility the capacity to activate the fecundation cycle which allowed the very evolution of the world, in the sense of assuring the continuity of the vital flow in the Andean cosmology.
In NC (c.243) the month is called *Camai Inca raymi quilla*, the moon of rest and of the feast of the Inca. The text reports that the Inca ruler held celebrations for the entire population, during which the ears of the nobles were pierced. Once again we see that there is concordance between the two sources, but the *Pachaquipu* emphasizes the symbolic relevance and content of the Inca rituals, starting from the very same name attributed to the month, which is more detailed as compared to the generic name given in NC.

[12° pendant] *aymuraypacha* = time for harvesting.

The cartouche contains the ticcisimi *sara*, corn. In the NC (cc.244, 245) we find a perfect concordance since the month is called *Aimoray quilla,* the moon of harvesting.

## 3. The astronomical content of the *Pachaquipu*

The knot marked black is the simple key which allows us to "decode" the data, because the *Pachaquipu* is declared by the author as a calendar which "concerns the end of the time of the Tahuantinsuyu": since the destiny of the empire of the Incas was decided the day of the battle of Cajamarca, and the Pachaquipu contains a single day marked with a black node, it is natural to assume that it represents November 16th, 1532, the day of the battle.[11] We therefore assume as a working hypothesis that the other knots of the Pachaquipu refer to the days before and after this date and proceed to control the correspondence of the data with the lunar and solar cycles, as well as with the other astronomical events reported.
Of course, the first problem to be solved is to individuate the latitude of the astronomer who compiled the calendar. It is natural to assume this latitude as that of the capital of the Inca, Cusco, and this is the second working hypothesis we shall use. Counting the number of the knots back and forth from November 16, it is immediately clear that the 365 days reported in the calendar start from June 3th, 1532 and arrive to June 2, 1533. The count of the lunar months agrees with the calculated phases of the moon in that period, starting from the new moon of June 3, and the identification of the solstices and the equinoxes is also coherent.
Hence the other symbols of the calendar reflect the astronomical events that occurred in Cusco in that period fairly well. In particular:

1) The eclipses.

To study the eclipses we use the tables provided by Fred Espenak, NASA's GSFC. The *Ticcisimi* representing the sun eclipse is not used in the calendar, and indeed no sun eclipse was visible, not only from Cusco but from the Inca empire as a whole, in the considered period of time. The *Ticcisimi* representing the moon eclipse instead on pendant 9[th] appears once, placed on February 9, 1533. Effectively, within the considered period of time a visible moon eclipse occurred that day. It was a very impressive total eclipse that reached its maximum just before dawn, around 7.10 a.m. (Cusco local time).

2) The Pleiades

The symbol named *Collcacapac* is a rectangle containing a shining star with eight jags, a sort of generic picture for "star" which occurs also in the Poma de Ayala's chronicle, where it is used for Venus. In any case there is no doubt that the asterism named *collca* coincides with the Pleiades.[12] Interest for the Pleiades in the Andean astronomy is indeed well attested, and clearly confirmed by the ethno astronomical study of contemporary Misminay carried out by Gary Urton in the eighties.[13] Among this people many Inca celestial concepts have survived, so that their information has been fundamental to understand important aspects of Inca astronomy which were described in a quite unclear way by the chroniclers, especially the so-called dark cloud constellations.[14]
In the Pachaquipu the symbol occurs two times, and it is located on April 5 and June 7. Clearly, the hypothesis to be tested is if these dates may indicate the period of invisibility of the Pleiades. As is well known, there is no rigid definition of the period of invisibility, since the observation of heliacal phenomena depends on the skill of the observer, as well as on the atmospheric conditions and, if the case, on the presence of a non-flat horizon at the azimuth of rising. In what follows we assume a flat horizon and use a reasonable estimate (-17°) for the altitude of the sun at which a star (actually in this case a dense group of stars) with an equivalent magnitude ~3 is firstly visible.[15] Assuming moreover a minimal altitude of visibility ~3° ("Thom's law", that is minimal altitude of visibility = magnitude in degrees) we get a theoretical period of invisibility at the latitude of Cusco in 1533 lasting from April 7-8 to May 25-26. The agreement with the *Pachaquipu*, which gives April 5 and June 7, is therefore optimal for the disappearance from the sky, while June 7 is too late to indicate heliacal rising. Perhaps it is an indication of a date in which the Pleiades, known to have returned in the pre-dawn sky from a few days, were actually looked at, at a sufficient altitude to distinguish their "seven sisters". In this connection it can be noticed that the chronicler Arriaga mentions the observance of the Pleiades in association with the maize production; the pre-conquest Inca feast lasted a few days and, after the conquest, was linked to the Christian feast of Corpus Christi, which occurs the ninth Sunday following Easter; in 1533 it occurred on June 15. Today, the heliacal rising of the Pleiades in the zone of Cusco occurs around June 3-4, but Urton reports that the observations - aimed to make predictions about planting - are carried out by the Misminay people on the day of St. John, June 24.[16]

3) The zenith passages of the sun.

As mentioned in the previous section, the *Pachaquipu* contains two yellow C-shaped cloth bands, whose meaning is not explained by Valera. However, due to their yellow color, they definitively refer to the sun. These symbols are placed in correspondence to the following days:

- the first starts between October 12 and October 13, and ends between October 17 and October 18, for a total of five days

- the second starts between February 5 and February 6, and ends between February 11 and February 12, for a total of six days

The dates October 19 - the day after the end of first band - and February 4 - the day before the beginning of the second band - are symmetrical with respect to the summer solstice (December 12 Julian). Thus they are, of course, suspicious of indicating solar dates, and it can indeed be verified that these two dates correspond to the passage of the sun at the zenith at the latitude of Cusco. Therefore, there can be little doubt on the fact that the C-shaped bands are used as zenith passages indicators.

## 4. The relationship between the *Pachaquipu* and the calendars in the *Nueva Coronica*

Since the two documents EI and NC were probably written by the same author, it is important to compare the EI *Pachaquipu* with the two calendars which appear in the *Nueva Coronica* (Fig. 3 and 4) We begin with the relationship between the Pachaquipu and the most well known of the NC calendars, which will be referred to here as the *calendar of the feasts* (to this calendar we have already made reference in Sec.2).
In comparing the names of the months in the two sources, we have already put in evidence that the two calendars are in general accordance. More precisely, six months share a perfect coincidence in their names, while the other six are called in a different way but their description in NC conceptually confirms the name given in the Pachaquipu. Interestingly, there is also accordance between the two sources on the anomalies respect to other calendrical sources: for instance, his holds for the names of the months and the ceremonies described for the $6^{th}$ (*Ayamarca*) and the $9^{th}$ (*Huarachicuy*) month, a fact which gives more credit to the idea that the two calendars arise from the same source. In order to understand the differences in the two calendars we must remember once more that NC was addressed to the King of Spain; it was therefore meant to be read by a Spanish audience, while EI, and its annexes written by traditional writing like the *Pachaquipu*, is directed to the native world and to the colonial descendants of the Inca Huayna Capac-Paullu Inca. As a consequence, the NC calendar of feasts presents a picture of the Andean concepts in a manner which is as much as possible understandable and *acceptable* to the western world, while the *Pachaquipu* was conceived to be more close to the Inca calendrical concepts: the two calendars thus form one *corpus* which give us some light on the Andean concept of calendar on the beginning of $17^{th}$ century (when both documents were written) and on the Inca concept of calendar as well. Indeed:
1) The *Pachaquipu* starts at the new moon before the winter solstice and adds the $13^{th}$ pendant with the 10 additional days in order to get a year of 365 days, while the NC *calendar of feasts* begins with the lunar month which follows the summer solstice. Curiously, however, this month is called "moon of the great feast of the sun" also if the solstice actually belongs to the previous month. Thus this month "looks like" our January, the first of the western month: the intention of realizing a compromise and an analogy with the great Christmas feast held at the end of December and the subsequent beginning of the western year is therefore pretty clear.

2) The *Pachaquipu* contains the word *pacha* in his name and in the names of all the months. The quechua word *pacha* means "time", but it means also "space", to stress that the calendar allows to count the time "focused" in the Andean cosmology, which is divided in the two opposites: *hanan* (the sky with the stars, i.e. the *hananpacha*) and *hurin* (the anthropized land, i.e. the *Pachamama*). The days of the year are divided according to periods of 10 days and of 15 days, in accordance with *hanan* (red) and *hurin* (green) alternance, as well as with the sun (hanan) and the moon (hurin). NC calendar uses instead the attribute *quilla* (moon) to describe the months; the focus is thus not in the complex sacred space, the *pacha*, but rather on the passing of time as it is seen from the anthropized earth, *Pachamama*. Therefore, the months are only divided in weeks of 10 days (NC, 260), where ten is the basis number of any Inca organization of the lands; further, the rituals are scheduled within the sacred *huacas* which organize the anthropized earth. [17] The unique case in which the term *pacha* appears among the names of the NC months is in March, the month of the vernal equinox (10th pendant of the *Pachaquipu* called *pacha pucuy quilla* in NC , that is ``month (moon) of the time of growing ``); it is tempting to suppose that this was a way of smuggling the cosmological importance of the vernal equinox, a moment at which the sun (*hanan*) and the moon (*hurin* - thought of as representing the dark hours) were equilibrated having the same "strength". This was a special, favorable moment, for entering into the cosmologic time, the *pacha*, i.e. to communicate with the Hananpacha. This is also suggested by the fact that sacrifices to Viracocha are described in NC as carried out during this month.

3) The presence of the *ticcisimi*, inserted in the *Pachaquipu*, mean that the time - the *pacha* - which was recorded on that kind of Quipu was not an abstract matter but, in a sense, a *tangible* one made up of many sacred elements. It was continuously flowing and rich of active - albeit potentially dangerous - sacred power, as in EI are said to be the *ticcisimi* by themselves. Indeed, according to EI a *Ticcisimi* "becomes active" in the *Hananpacha* whenever it is tight to a sacred quipu; the *ticcisimi* have therefore to be "caught" by the priests, in order to "fix" their dangerous and fluid elements in the anthropized earth and, in this way, transform them into *huacas*, readable by the *tocapus*. What is of special interest here is the connection of this symbolic mechanism with the relationship between astronomy and power which, as is well known, is present in almost all cultures.[18] Indeed, we notice the existence of *ticcisimi* having an indubitable astronomical content, such as those related to the sun or to the Pleiades. As said, an "active" *ticcisimi* of this kind was supposed to be "wandering" until the priest - in this case the *amauta* - was able to "catch" it and fix it in a huaca. "Catching" an "astronomical *ticcisimi*" clearly meant making the duties of the astronomer related to the observation of the corresponding celestial cycle, and it is actually very well known that several huacas of the Cusco ceque system were connected with astronomical observations.[19] On the contrary, the time of the NC calendar is presented as "already fixed in the huacas", since NC describes the ceremonies held every month in the anthropized earth: i.e the NC time is a "linear" time acting where men lives and acts. Once more, we see that the two calendars are complementary.

There is also one curious aspect of the complementary of the two calendars, which is connected to those astronomical phenomena mentioned in the *Pachaquipu* which are in danger of being considered idolater's ceremonies. This occurs more clearly when the corresponding observations were absent from western astronomy and, consequently, unrelated to those Christian festivities which inherited an astronomical connection. This is true, in particular, for the Zenith passages, which of course do not occur at European latitudes. As a consequence, we do not find any mention of the zenith passages of the sun in NC while the Pachaquipu reports them (with the two "C" symbols discussed above) but *without explanation*. Was it just an oversight? Actually the passages of the sun to the zenith must have been related to specific ceremonies among the Incas. The 5th month, corresponding to the first passage, should have been dedicated to the *Spiritus Vitalis* (one of the main forces in the Hananpacha, see Sec. 3). Similarly, the second passage of the sun at the zenith in the 9th month should have been accompanied by "prohibited rituals", namely those mentioned as such in NC: *uarachico* and *rutochico,* which deal with another very important force in the Hananpacha, the power of masculinity named by EI (add.III) also as *Amaru,* the destructor. There is, therefore, a sort of *underlying hierarchy* between those things which are considered as fully acceptable also by the King (such as the December solstice/ beginning of the year festivals and the feast of the deads) and are therefore fully discussed in NC and explained at the Inca manner in the Pachaquipu, others which are at strong danger of idolatry (such as the initiation and earrings rituals of the months 9th and 11th ) which one finds mentioned in NC but in a subdued way, i.e. deprived of the evidence of the corresponding month name, but are re-evaluated in the Pachaquipu. Finally in this trail of prudent hiding there are two typical astronomical/religious Inca sky-watching rituals, the Zenith passages of the sun on the 5$^{th}$ and the 9$^{th}$ month, and also the heliacal setting/rising of the Pleiades, which we see in the *Pachaquipu* while the Zenith rituals are absent in the NC description as well as in the pictures of the corresponding February and October months, and apparently no cult to the Pleiades is mentioned in the NC calendar. Actually, the cult of the Inca to the Pleiades is mentioned cryptically, thus only for Natives, by the *tocapus* painted on the Incas' belt in the months of February, April, June, October and December subduing the cosmic relation between the Pleiades and Inti, the Sun as well as with the Inca, the Sun's son, who is wearing their tocapus: the Pleiades which, according to EI (c.11r), were "the Sun' nourishment" and thus are thought of as sustaining the Inca in making efforts.

As mentioned above the NC contains also a second calendar, located at the end of the document (cc.1130-1167). The text accompanying this second calendar begins with some observations on the "hard" climate of Castilla and the mild climate of the Andean region, and, consequently, puts in evidence the better conditions of life existing in the latter region: the poverty of the Andean people is consequently attributed only to the oppression on the Indios. The calendar looks like an Andean version of an European farmer's calendar of the 17$^{th}$ century, and this is why we name it *Agricultural Calendar*, but it is built in parallel with the above mentioned calendar of feasts: it contains indeed a descriptive picture for each month under the title of the month which is instead named "duty" and the indication of the agricultural duty of that specific month (for instance July is *duty: pack away the corn and the potatoes*) followed by the description of the duty and a table with the name of the saints of each day of the month, divided in 7-days weeks. Hence the names of the months are completely different from those mentioned in the Calendar of Feasts (these names are however reported as well, in small letters). The pictures of each month show farmers, instead of Incas, and the moon is portrayed infrequently, while the sun is almost always present: hence we consider it a solar calendar. The unique drawing which coincides with that of the previous calendar is that of August, the month of plowing here named "*duty: triumphal chants, time to open up the earth*". In both pictures of the two calendars of NC indeed we see the Inca governing the activities, and the unique sensible difference is that the people of his group are looking right (i.e. in accordance with the EI direction of *hananpacha*) instead of left, which is the direction associated with *hurinpacha* of the *calendar of feasts*. The simultaneous presence of both calendars in NC suggests, on one side, an attempt to propose analogies with the Europeans, showing one calendar for the religious feasts and another for agriculture, and on the other side a way to bequeath to the Natives some of the cosmogonyc traditional moments not to be forgotten.

All in all, from this comparison a clear convergence and a sort of complementarily between the two sources arises, which is in accordance with the idea that they were both written by the same hand or at least that they aroused within the very same cultural group.

## 5. The *Pachaquipu* and the Inca astronomy

As we have seen, the *Pachaquipu* is a lunar/solar calendar of 365 days containing November 16, 1532, starting from the new moon of June 3, 1532, valid at the latitude of Cusco. It is out of question the fact that the document is original: it was certainly drawn by Blas Valera at the beginning of the 17$^{th}$ century. However, in a sense, the fundamental problem of "originality" starts right here: it is the problem of understanding how much information about pre-conquest astronomy is contained in it, and how much instead is inherited from the author's knowledge of western astronomy.

Of course, a sort of tautological difficulty immediately arises since our current knowledge of pre-conquest astronomy is fragmentary and many points are still widely debated among specialists. We will, therefore, try to refer only to those pieces of information which appear to have been well established.[20]

The first issue which has to be investigated is, if the calendar structure of the Pachaquipu is in agreement with the reports of the chronicles about the Inca calendar, as well as with known pre-conquest calendars. From the chronicles it is indeed well attested that the Incas had a lunar calendar, in which each month started with the new moon, as well as a solar calendar of 12 months of 30 days with 5 "epagomenal" days.[21] Many details are however unclear. In particular, it is unclear if – as it has been suggested – the re-alignment of the lunar calendar with the solar one occurred one time each three years. We do not have enough information on this point, and actually the Pachaquipu adds all the ten days needed in a sort of ``epagomenal month``. Since the count in 15 days completes at the 360$^{th}$ day, the ten added days actually correspond to five red knots and five green knots; due to the fact that, as we have seen, the alternation of colors correspond to the *hanan/hurin* dualism, this might be interpreted as a sort of indication that the epagomenal ``month`` was perhaps added *each* year.

Another point which is quite unclear, because the different Chroniclers do not agree each other, is the beginning of the year,[22] since Molina and Diego Fernandez begin the year with the lunar month which includes the winter solstice but the Anonymous Chronicler mentions the month of the march equinox and finally Betanzos and, as we have seen, the NC chronicle indicate that of the summer solstice. From this point of view, the Pachaquipu supports the view that the original Inca year started at the winter solstice, and its ``companion`` NC turned it to the month corresponding to January in order to comply with the western standards.

So far for the information provided by the chroniclers. As far as the relationship of the Pachaquipu with available *pre-conquest* documents is concerned, an analysis has already been carried out by Tom Zuidema.[23] In his work Zuidema compares the Pachaquipu with two surviving Quipu calendars - the first from Ica, made in Inca times, and the second from Chachapoyas made in late Inca times or early colonial years - as well as with a calendar textile from earlier (Huari) times. Both the calendars which are certainly pre-conquest contain relevant common points with the Pachaquipu; indeed the Huari calendar counts 12 months of 30 days plus five extra days and thus corresponds to the "solar calendar" of the Pachaquipu, although it makes no reference to lunar months, while the Ica Quipu calendar combines a solar with a lunar count in a way which is similar to that of the Pachaquipu (the Chachapoyas calendar instead describes months of either 30 or 31 days with one of 29 days, and therefore has been almost certainly influenced by the European calendar). Zuidema's conclusions are that Blas Valera "in some way had access to the Inca calendar"; according to this author perhaps Valera studied the problem of the calendar when he was working for the 3rd Concilio Limense, held in 1582-1583, although in the acts of the Concilio there is no mention of a calendar with a lunar count.

We can conclude that Valera had access to Inca informants in the second half of the 16$^{th}$ century, and that the Pachaquipu, although coherent with, does not give new information on the structure of the Inca calendar. Its contribution from this point of view is rather of ethnological anthropological type, since - as we have seen in the list of the months - it allows a better understanding of the cycle of the Inca rituals in the course of the year. It is actually possible to re-read the *Pachaquipu* together with the reconstruction of the Inca calendar and festivals for the years 1532-1533, with the aim of making precise as much as possible the schedule of the Inca festivities in Cusco, as will be shown in a future work.

So much for the general calendrical structure; however, what about the *data* contained in the document? Indeed, the Pachaquipu contains specific data of the years it concerns, such as the new moon on June 3, 1532, and the total moon eclipse on February 9, 1533; further, it contains data which are typical of the latitude of Cusco, such as the dates of the Pleiades disappearance/appearance and the zenith passages of the sun. There are here two, quite different possibilities. It may be, in fact, that Valera compiled the calendar by himself. In this case he worked it out in accordance with the Inca customs, but evaluating the data of 1532-1533 by himself or with the help of an expert brother. Indeed, the Jesuits developed a deep interest in sciences in general and in astronomy in particular since their foundation in 1540.[24] They founded dozens of observatories, and relevant astronomical observations by Jesuit astronomers are documented already at the end of the 16$^{th}$ century, for instance by Fr. Cristopher Clavius, active in Rome the *Collegio Romano* observatory, and by Fr. Cristopher Sheiner (1573-1650) who, among other contributions, was the discoverer of the sun-spots. Therefore, it is certainly possible that Blas Valera had access to astronomical data and/or to skilled colleagues, not to say that we do not know how deep was his own knowledge of astronomy. Access to "almanacs" (databases) containing eclipse's tables was possible as well, as shown for instance by the famous episode of the eclipse predicted by Columbus on February 29, 1504 and used by him to impress the natives.

The second, quite fascinating possibility, is that Valera collected not only information on the general structure of the Inca calendar, but also a set of *original data* recorded by Inca astronomers. This is a quite interesting scenario for the following reason. The existence of specialized Inca astronomers is out of question, and it is very likely that they used Quipus to store the data of their observations; actually the NC explicitly represents one of them with a fork-like viewfinder (a somewhat classical instruments, useful to increase naked-eye accuracy, used by many skywatchers throughout the history, from the Egyptians to the Mayas) and a Quipu (Fig. 5). However, among the six hundreds or so of quipus which survived the conquest, only a few are calendrical, and not even one has been as yet identified as a storage of astronomical data.[25] Therefore, the *Pachaquipu* would constitute the first - although quite indirect - proof of the existence of such astronomical Quipus. Hints in this direction come especially from the references of the *Pachaquipu* to the zenith passages of the sun. Indeed, of course, this phenomenon was not experienced by European astronomers in their countries, and therefore it was known essentially at a *theoretical* level. As a matter of fact, the Inca observations of the zenith passages of the sun should have generated much confusion in the chroniclers, as it is shown by the ambiguous description made by Garcilaso de La Vega, which is only valid near the equator, since it associates the zenith passages - measured trough the shadow cast by a gnomon - with the equinoxes. On the other end, many chroniclers report the existence, on the hills of the Cusco horizon, of stone pillars (*mojones*) used as calendrical indications for sunset at specific days, and the Anonymous Chronicler describes the site of the observation as the *ushnu*, a pillar of well-worked stone said to have been located in the today's Plaza de Armas, not far from the main Cusco temple. Unfortunately, no traces of the stone pillars have been found as yet in Cusco, and it is unclear which dates they really defined; however, their former existence is out of question, since examples are known from other Inca sites such as Urubamba and, recently, the existence of a long-standing tradition of sun observation using towers at the horizon has been spectacularly confirmed by the discovery of the pre-Inca observatory of Chankillo.[26] Further, it is almost certain that the Cusco pillars were working as *coupled devices*; in other words, they were all built in nearby couples, so that the setting sun could be framed between two pillars for a period of time of the order of some days, helping the astronomer to identify the dates with precision.[27]

All in all, it is tempting to suppose that the "C" shaped symbols may represent the interval of days in which the sun was seen as framed between two pillars at the horizon; perhaps Valera copied the symbols without even being aware of their *practical* meaning and for this reason he did not mention them in the list or, as we have already proposed, he choose not to mention explicitly an observation which was carried out with devices which might have been regarded, due to the Inca cult of the sun, as "idols" located at the Cusco horizon.

## 6. Discussion and conclusions

The EI document has a rigorous logical structure, where nothing seems to be due to a chance. Adding the Pachaquipu to this document must have been a well pondered decision, and it is clear that Valera considered it as a fundamental piece of information about *both* the "old time" of the former *Tahuantinsuyu*, depicted - in a sense *frozen* - in the year of the fatal events of 1532, as well as about the (hoped) "new time" of the new Inca-Christian Kingdom. The document has therefore two different functions: it is an actual calendar of 1532-1533 and it also meant as an example of a calendrical *scheme* to be used in the Inca-Christian Kingdom.

From the point of view of the "old time", the Pachaquipu is, as a matter of fact, the unique documentation found so far about the quipus used to store astronomical data which, as we have seen, almost certainly existed. Among other information, it gives us the first explicit example of the registration of the period of invisibility of the Pleiades in Cusco, which is reported to be of 61 days. This is too long for an expert skywatcher (in particular, the date of heliacal rising is too late) but is in any case very far from the Zuidema[28] proposal of a period of only 37 days, which in turn would complete a "ritual" ceques calendar connected with the sidereal motion of the moon and the number of huacas of the ceque system, which is 365-37=328. Therefore, the present document does not support the existence among the Incas of such a lunar-sidereal calendar. The lunar calendar of the Pachaquipu is instead a sinodic one, as reported by many sources about the pre-conquest Inca calendar; further, there is no evidence that the "epagomenal" days were accumulated each three years in an additional month, as it has been sometimes suggested. Instead, the calendar deals with the additional days needed to harmonize the lunar/solar count in a very peculiar manner. Indeed, these days are put (in the 13th pendant) under the protection of the sun, since the cartouche is a yellow square. It follows that the string "belongs" to *Inti* , the Sun, represented on the earth by the Inca. Symbolically, this suggests that the capture of the Inca in 1532 not only caused the fall of the empire but also altered the cosmic order, to be restored by the sought-for arrival of a new Inca. The interplay between red and green colors in the calendar days recalls the duality of opposites, *hanan* and *hurin*, which in turn are related to the Sun, the stars and "the world over" and to Pachamama, the Moon, and the "actual" world (e.g. farming, taxes and economy) respectively. However, these two fields are not so clearly divided as they would be according to the European "linear" way of thinking. They are indeed interrelated each other in accordance with the so-call ed *holistic* viewpoint that characterizes the Andean world,[29] as the distribution of the festivals in *hanan* and *hurin* days and the very same existence of different names for the months in the two documents EI and NC confirm.[30]

To this picture we should certainly add the companion issue of understanding the syncretistic views of the author and of his group about the *new* time and space, the new reign to be realized in Paititi. Of course a deep analysis of this problem would go beyond the scope of the present paper, however from the point of view of the History of Astronomy and Calendrics we can at least notice that, in a sense, the Pachaquipu represents the merging of the two calendars which the same author presents in the NC: the first is indeed essentially of the lunar type, the second mainly solar, while the considerable complexity of the Pachaquipu is due to the fact that the lunar and the solar counts run together. Probably, at least in our opinion, although the division between the lunar count associated with the festivities ("sacred calendar") and the solar count associated with practical activities ("agricultural calendar") proposed in the NC is prepared for the Spanish mentality, the simultaneous presence of both calendars in NC, together with the image of the Inca performing the same ceremony of plowing in the calendar of feasts *but seen from the hanan word* leads to think that both NC calendars were complementary also for the Natives. The aim was therefore not only to recall the lunar/solar calendar which was in use in pre-conquest times divided into two parts, one (*hurin*) used to schedule the festivities and the other (*hanan*) used to organize the work in the fields, but also that of bringing, in secret, some of the cosmogonyc traditional festivities to be held in Paititi, where probably both calendars had to be used. However, the pre-Colombian Inca calendar should have been much more similar to the Pachaquipu, without a neat division between "sacred" and "agricultural" cycles, since this division did not exist among pre-Columbian Incas. As a consequence, this document furnishes us a picture which appears to contain, although admittedly obfuscated and mediated, at least some of the elements of the original structure of the *pacha*, the sacred space and time of the Incas. In addition, it suggests an interesting interpretation of the role of the Inca amauta astronomer in terms of Archaeology of Power. Indeed, it appears that his role was to carry out the astronomical observations related to calendar days associated to powerful but potentially dangerous cosmic forces, in order to calculate and transferring them, by the abacus or *yupana*, on the earth in form of venerable but potentially controllable *huacas*.

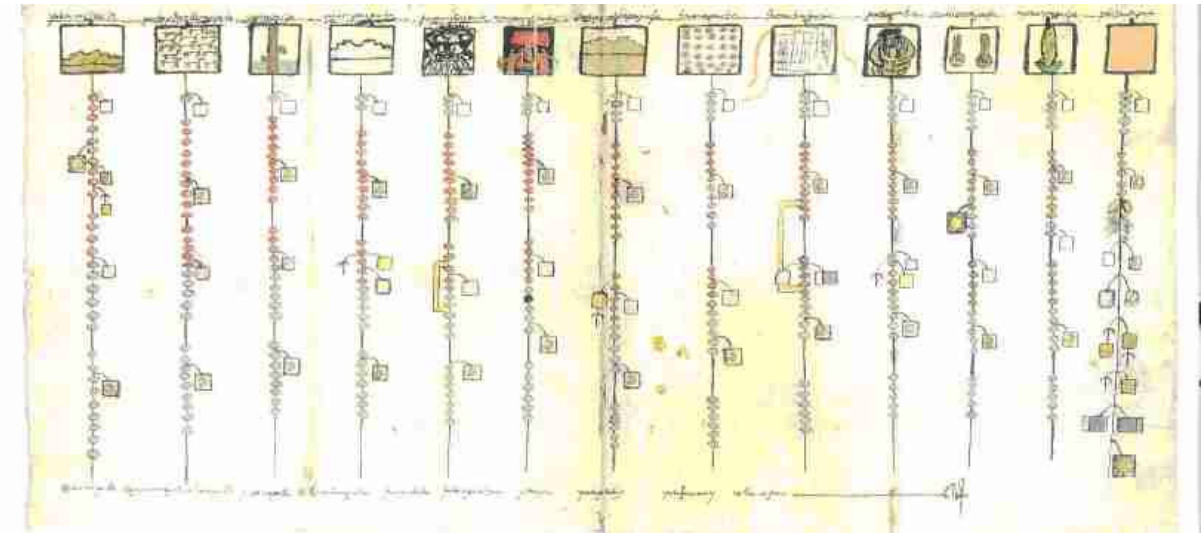

Fig. 1. The *Pachaquipu* (Miccinelli Collection, copying prohibited, published with the kind permission of Clara Miccinelli)

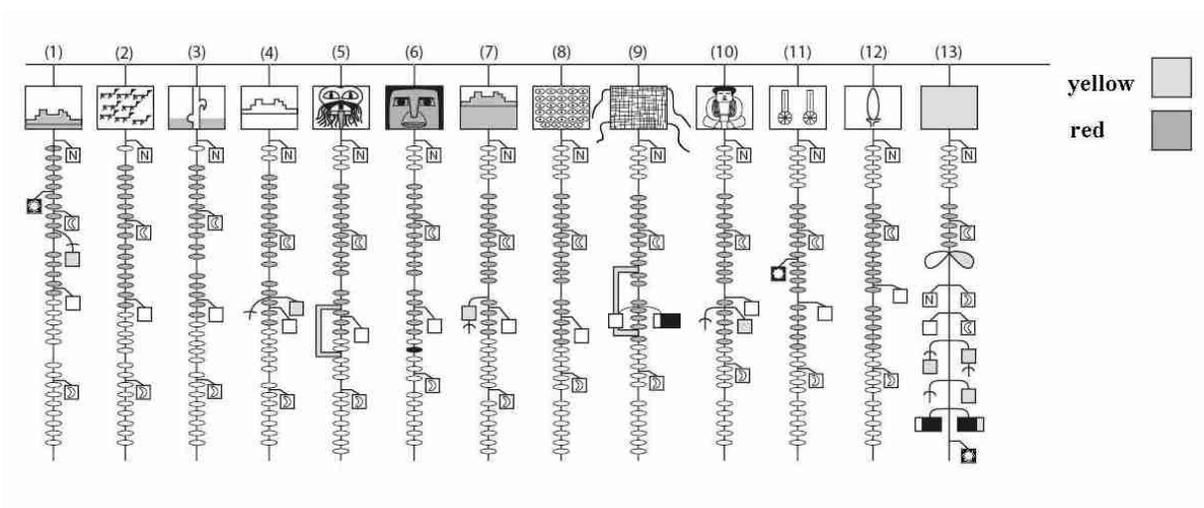

Fig. 2. Graphical scheme of the *Pachaquipu*

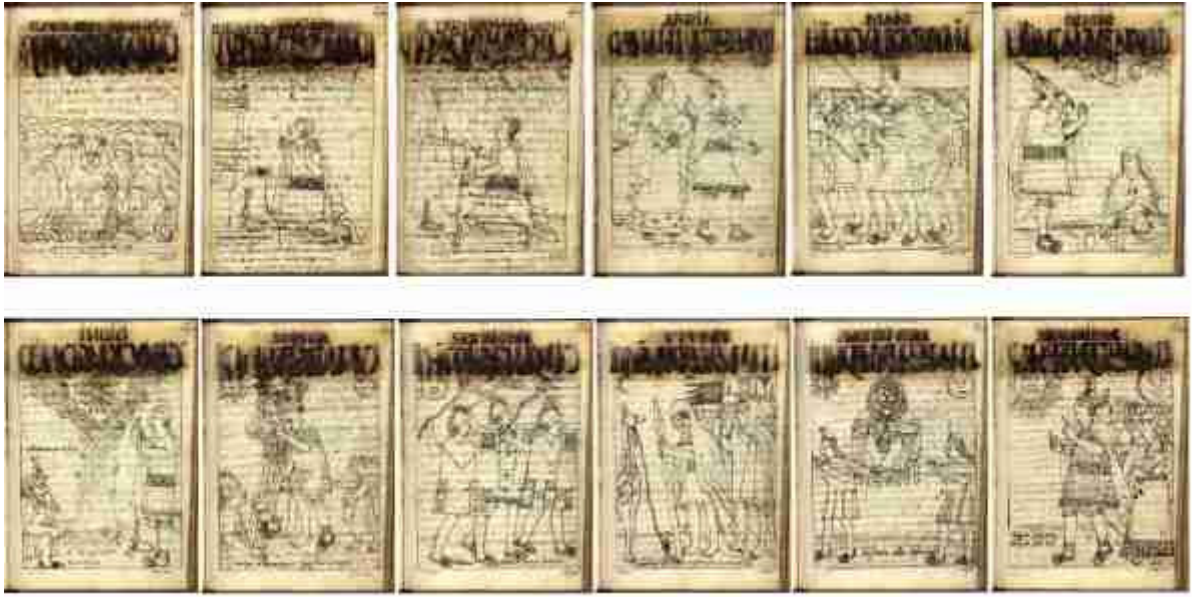

Fig. 3 The pages reporting the images of the "Calendar of the feasts" in NC

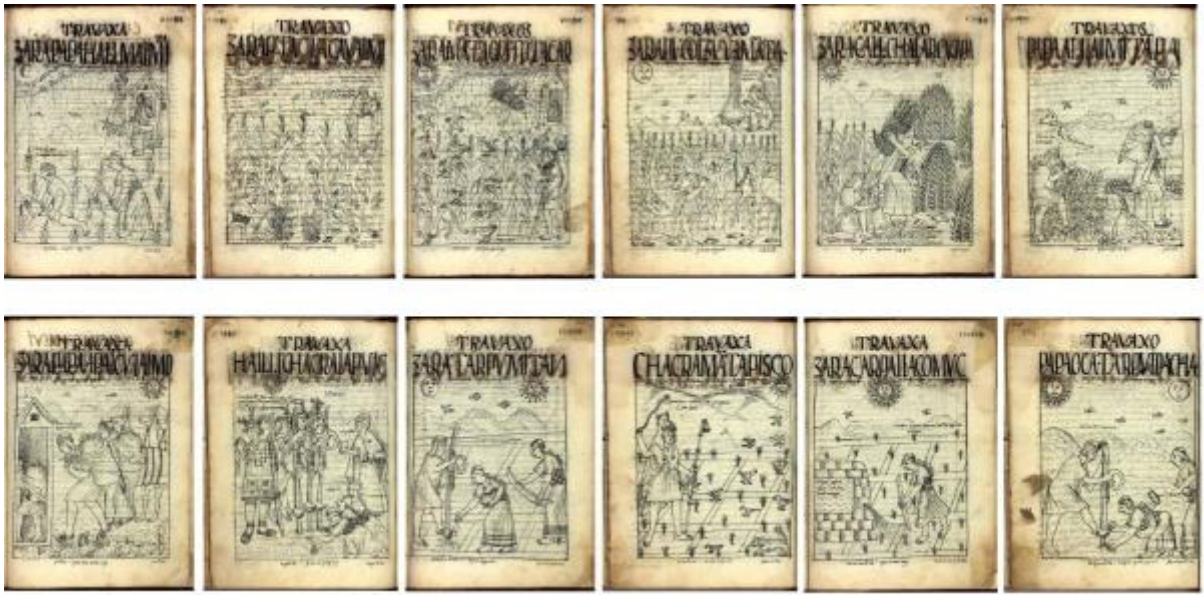

Fig. 4 The pages reporting the images of the "Agricultural calendar" in NC. Notice that only the 8$^{th}$ month is represented with the same picture of the other calendar, although in "inverted direction".

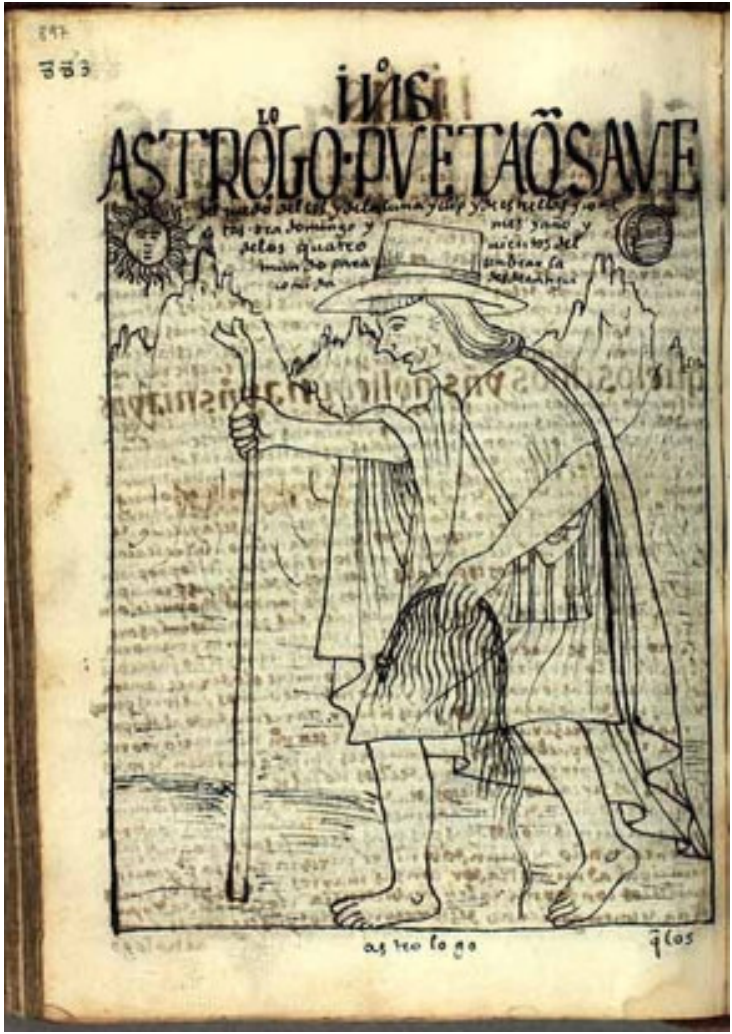

Fig. 5 The Inca "astrologer", carrying a Quipu and a fork-like post, in the NC

**References**

[1] Laurencich Minelli, L. ed. (2007) *Exsul Immeritus Blas Valera Populo Suo e Historia et Rudimenta Linguae Piruanorum. Indios, gesuiti e spagnoli in due documenti segreti sul Perù del XVII secolo*. CLUEB Bologna. From now on *cartae* and *addenda* of these two documents are cited in accordance with this transcription.

[2] Authenticity has been proved by several technical analysis as well as by the examination of related documents of the same period. Among the technical probes we mention the analysis of colours, inks, paper sheets and graphology carried out mostly at the Bologna University, and the application of the Carbon Dating technique carried out by the Australian Nuclear Science and technology Organization, Menai, Australia. All these analyses agree in showing that the document was written at the beginning of XVII century by the same hand who signed it: the Jesuit Blas Valera. Among the "cross" documents which confirm the facts mentioned in EI there are manuscripts in public archives; in particular, in the *Archivio di Stato di Napoli* two letters that Francisco de Boan wrote from Lima to the Napoli viceroy Conde di Lemos (1610, 1611) are

preserved. In these letters the writer informs that the Jesuit Blas Valera was alive and back in Perù. Another relevant documents has been found in the ARSI Archive in Rome, signed Blas Valera and dated Alcalà, 25 de Junio 1618. It is addressed to F. General Muzio Vitelleschi to let him know that Exsul Immeritus was ready to be sent to him. The matter is in continue evolution because additional details and cross-references are being found in different public archives; for full details on technical and historical proofs we refer the reader to Francesca Cantù (ed.), *Guaman Poma y Blas Valera. Tradicion Andina e Historia Colonial.* Pellicani, Roma (2001), 143-194, and to the following proceedings of conferences: Laurencich-Minelli Laura and Piero Mannucci (eds.) *Atti del convegno internazionale: nuove prospettive negli studi andini*, Archivio per l'Antropologia e la Etnologia. CXXXV, Firenze, (2005); Laurencich-Minelli Laura y Paulina Numhauser (eds) *El Silencio protagonista. El primer siglo jesuita en el Virreinato del Perù. 1567-1667.* Actas del Simposio Int. Hist.11, 51° Congreso Internacional de Americanistas, Santiago de Chile, 16-20 de julio 2003, ABYA AYALA, Quito; P. Numhauser e M. Casado Arbonies ed. (2006) *Escrituras Silenciadas en la Epoca de Cervantes,* Universidad de Alcalà, Alcalà de Henares; L.Laurencich Minelli e D. Domenici, eds. *Per bocca d' altri: Inca, gesuiti e spagnoli nel Perù del XVII secolo.* Alma Digital Library, Università di Bologna, 2007, (http://almadl.cib.unibo.it); Laurencich-Minelli Laura y Paulina Numhauser (eds) (2007) *Sublevando el Virreinato. Los documentos Miccinelli como documentos contestatarios a la historiografía tradicional del Perú colonial.* Actas del Simposio Int. Hist. 39, 52° Congreso Internacional de Amerricanistas, Sevilla, 17-21 de julio 2006, ABYA AYALA, Quito.

[3] The *Capacquipu* are syllabic quipus used in EI for sacred chants. The writings are obtained by subsequently inserting the sacred symbols or *ticcisimi*, each one with a cord. The number of knots in the cord serves to indicate which syllable has to be extracted from the name of the symbol (for instance a two-knots cord under the ticcisimi *Pachacamac* indicates the syllable *cha*). The *yupanas* are abacus, used to transform the words in sacred numbers, while sums of such numbers are reported in standard numerical quipus. See Laurencich-Minelli L., (1996) *La scrittura dell' antico Perù*, CLUEB, Bologna; Id. (2004), *Quipu y escritura en las fuentes jesuiticas en el virreinato del Perù entre el final del siglo XVI y la primera mitad del siglo XVII*: En L. Laurencich-Minelli y P. Numhauser ed.: *El Silencio protagonista. El primer siglo jesuita en el Virreinato del Perù. 1567-1667.* Abya Ayala, Quito, Ecuador, 2004: 171-212; id. (2007) *Presentacion de los documentos Miccinelli in Laurencich Minelli*, L. ed.(2007) *Exsul Immeritus Blas Valera Populo Suo e Historia et Rudimenta Linguae Piruanorum. Indios, gesuiti e spagnoli in due documenti segreti sul Perù del XVII secolo:* 103-184. CLUEB Bologna.

[4] Magli, G. (2005) *Mathematics, Astronomy and Sacred Landscape in the Inka Heartland* Nexus Network Journal - Architecture and Mathematics 7 p. 22-32; Laurencich-Minelli L. (2001) *Il linguaggio dei numeri, dei fili e della musica presso gli Inca*, Esculapio, Bologna.

[5] Laurencich-Minelli 2006 *Il ragno nelle antiche culture andine: un tratteggio*. "Quaderni di Thule" *2005*, Perugia: 45-56.

[6] Buongiorno, V. (2007) *La lengua quechua en los documentos Miccinelli*, In: Laurencich Minelli L.(ed), (Ref. 1); Laurencich-Minelli L. and Rossi, E. (2007), *La yupana de la nueva coronica y las yupanas de Exsul Immeritus Blas Valera Populo Suo: abaco y escritura inca o sincretismo jesuita?* In: Laurencich-Minelli L. and P. Numhauser ed, *Sublevando el Virreinato*, Abya Ayala, Quito.

[7] Valcarcel, L. E., (1965) [1949], *Ruta cultural del Perù*, ed. Nuevo Mundo, Lima; Id. (1972) [1927] *Tempestad en los Andes*, Ed. Universo, Lima. For the development of the plan of Paititi see Laurencich-Minelli L. (2007) *Los documentos Miccinelli (siglo XVII): dos curiosos testigos sobre una utopica reduccion "Inca"/Jesuita en la Provincia Peruviana*, in Laurencich-Minelli L. and P. Numhauser, Ref.6.

[8] According to EI, *ticcisimi* is a textile cartouche with phonetic reading, i.e. it can be read only in the language used in that text written by the *capacquipu* (the quipu for writing texts, see Ref. 3) as well as by any other sacred quipu like the Pachaquipu; *tocapu* is a textile cartouche with

ideographic figures which has always a conceptual reading, i.e. it can be read in any language. We often find the same figure used as *ticcisimi* when tighted to *capacquipus* (as well as in the Pachquipu itself)*,* or used as a *tocapu* in other contexts.

[9] Laurencich-Minelli, Ref. 4 and Laurencich-Minelli and Rossi, Ref. 6.

[10] Zuidema (2004) *El quipu dibujado calendárico llamado pachaquipu en el documento Exsul Immeritus de la collección Miccinelli.* In El Silencio protagonista. El primer siglo jesuita en el Virreinato del Perù: 1567-1667:222. ABYA AYALA, Quito.

[11] We stress that all the dates in this paper are Julian.

[12] Bauer, B., and Dearborn, D. (1995). *Astronomy and Empire in the Ancient Andes*, University of Texas Press, Austin.

[13] Urton, G. (1982). *At the Crossroads of the Earth and the Sky: An Andean Cosmology*, University of Texas Press, Austin.

[14] See Magli, Ref. 3

[15] Aveni, A. F. (2001). *Skywatchers: A Revised and Updated Version of Skywatchers of Ancient Mexico*, University of Texas Press, Austin.

[16] See Bauer and Dearborn, Ref. 15; Urton, Ref. 4

[17] For the subdivision of the ritual calendar into two halves likely corresponding to *hanan* and *hurin* see Ziolkowski (1989) *El calendario metropolitano inca* and Ziolkowski (1987), *Las fiestas del calendario metropolitano Inca: primera parte.*

[18] For a comprensive account on this subject see Krupp, E. C. (1997). *Skywatchers, Shamans, and Kings*, Wiley, New York; Magli, G. *The power from the stars: an introduction to archaeoastronomy* Praxis Publ. in Astronomy and Astrophysics, at press.

[19] See Zuidema, R. T. (1964), *The Ceque System of Cusco: The Social Organization of the Capital of the Inca*, Brill, Leiden; Bauer, B. (1998), *The Sacred Landscape of the Inca: The Cusco Ceque System*, University of Texas Press, Austin.

[20] For a complete review and a summary of different viewpoints see Aveni, A.F. (2003) *Archaeoastronomy in the Ancient Americas* Journal of Archaeological Research, Vol. 11, No. 2, p.149, and references therein.

[21] The possible, although debated, existence of a further Inca calendar based on sidereal lunar months and connected with the so-called *ceque* system appears to have no recognizable bearing with the calendrical structure of the *Pachaquipu* and consequently is not discussed in details here; for the geo-quipu *cequecuna* of EI see Laurencich-Minelli (2004) *Ulteriori prospettive per la lettura dei quipu. Il quipu di Firenze n. 3887.* Archivio per l' Antropologia e la Etnologia*,* CXXXIV: 101-126, Firenze; and Laurencich-Minelli, *Presentacion de los documentos Miccinelli in Laurencich Minelli*, in Ref. 3.

[22] Bauer and Dearborn, Ref. 3.

[23] Zuidema, T. (2002) *Los Días Epacta y Epagóminos en Calendarios Pre-Hispánicos del Perú y Según Opiniones de Cronistas*. In *Il Sacro e Il Paesaggio Nell' America Indigena*, D. Domenici, C. Orsini and S. Venturosi eds. Bologna; Zuidema T. (2004) *El quipu dibujado calendárico llamado pachaquipu en el documento Exsul Immeritus de la collección Miccinelli*. In *El Silencio protagonista. El primer siglo jesuita en el Virreinato del Perù 1567-1667*: 172-212., L. Laurencich-Minelli & P. Numhauser ed. ABYA AYALA, Quito.

[24] Udías, A. (2003) *Searching the Heavens and the Earth: The History of Jesuit Observatories* Springer; NY.

[25] See Urton, G (2003) *Signs of the Inka Khipu: Binary Coding in the Andean Knotted-String Records* Un. Of Texas Press, Austin, and references therein.

[26] Ghezzi, I, and Ruggles, C. (2007) *Chankillo: A 2300-Year-Old Solar Observatory in Coastal Peru*. Science 315. no. 5816, pp. 1239 – 1243.

[27] A similar way of increasing accuracy in solar observations by means of a coupled device is probably embodied in the geometry of the windows of the so called *Torreon* of Machu Picchu, see

Dearborn, D. and White R., (1983) *The 'Torreon' of Machu Picchu as an Observatory*, Journ. Hist. Astronomy 5, pp. S37-S49.

[28] See Ref. 22, or Ref. 3 for an up-to-date discussion.

[29] Holistic logic is apparently absurd, since "the total is greater than the sum of the parts"; it is actually a process which transforms itself enlarging in a sort of "reflecting mirrors" interplay. This is particularly clear in the case of the sacred arithmetic discussed in EI, where the numbers associated to gods and to natural forces mix up to produce a divine unity, which is however different from the simple "sum of the parts". This holistic approach is present not only in the precolumbian Andean world, but also in Meso-America; see e.g.Laurencich-Minelli (2007) Ref. 3, Laurencich-Minelli (2004), *Lo zero concreto nel mondo inca e maya e cenni sul calcolo degli Inca.* In: *Calcolo precolombiano* Julio Macera Dall' Orso ed. Bardi, Roma, pp.289-314; - Id., (2003) *Quipu*, Ref. 21; Id, *Lo sagrado en el mundo inca después del III Concilio Limense de acuerdo a documentos de la época. Un esbozo*. In: *Simposio ARQ-8, Tawantinsuyu 2003: Avances recientes en arqueología y etnohistoria*. Coords. R. Barcena y R. Stehberg, *Xama, Publicación periódica de la Unidad de Antropología* 15/17. INCIHUSA, CRICYT: 241-254, Mendoza, 2002-2005. Id. (2005), Ref. 5; Id., (2006) *La yupana: abaco e ordinatore dello spazio Inca o esempio di sincretismo culturale?* "Quaderni di Thule" 2006, Perugia: 313-326, Id. op. cit. 2007, pp.144-167; Laurencich-Minelli and Rossi (2007) Ref. 6. ; Lopez-Austin A. (1983) *Nota sobre la fusión y la fisión de los dioses en el panteon mexica*. «Anales de Antropologia », XX, 2, 75-87. Holism is documented as well in todays'etnographical studies; for instance in the textile language of the Aymara - see e.g. Arnold, D. et al.(2000) *El rincon de las cabezas*, Colección Academia nueve, La Paz: 273-430. Here the Andean society is presented as a "textile" living being who has his/her own identity but it is also a "part of the whole". The same way of thinking corresponds to that described by Emmanuel Desveux (2001) (*Quadratura americana. Essai d' antropologie lévi-straussienne*. Georg, Collection Ethnos, Genève) among the today Ojibwa, the Iroques and the Sioux people, where mythology, rituals and sociology appear to have an interchangeable relationship each other; actually this way of thinking probably was a common substrate in the whole pre-Columbian world.

[30] Besides the NC and the EI documents, the fact that the names of the months were not unique is shown, for instance, by the early 17th century document "*Discurso de la sucesión de los Yngas*" by Anonymous; most authors refer to them as fixed because these names agree in three different 16th century sources (Juan de Betanzos, Polo de Ondegardo, Cristobal Molina) who however probably got their information from the same original.